\documentclass[aip,cha,amsmath,amssymb,reprint]{revtex4-1}

\usepackage{graphicx}
\usepackage{dcolumn}
\usepackage{bm}

\usepackage[utf8]{inputenc}
\usepackage[T1]{fontenc}
\usepackage{mathptmx}
\usepackage{etoolbox}
\usepackage{xcolor}

\newcommand{\R}{\mathbb{R}}
\newcommand{\dx}{\mathrm{d}}

\newcommand{\revise}[1]{{\color{black}#1}}

\makeatletter
\def\@email#1#2{%
 \endgroup
 \patchcmd{\titleblock@produce}
  {\frontmatter@RRAPformat}
  {\frontmatter@RRAPformat{\produce@RRAP{*#1\href{mailto:#2}{#2}}}\frontmatter@RRAPformat}
  {}{}
}%
\makeatother

\begin{document}

\title{Time-series-analysis-based detection of critical transitions in real-world non-autonomous systems}
\author{Klaus Lehnertz}
\email{klaus.lehnertz@ukbonn.de}
\affiliation{Department of Epileptology, University of Bonn Medical Centre, Venusberg Campus 1, 53127 Bonn, Germany}
\affiliation{Helmholtz Institute for Radiation and Nuclear Physics, University of Bonn, Nussallee 14--16, 53115 Bonn, Germany}
\affiliation{Interdisciplinary Center for Complex Systems, University of Bonn, Br{\"u}hler Stra\ss{}e 7, 53175 Bonn, Germany}

\date{\today}

\begin{abstract}
Real-world non-autonomous systems are open, out-of-equilibrium systems that evolve in and are driven by temporally varying environments.  
Such systems can show multiple timescale and transient dynamics together with transitions to very different and, at times, even disastrous dynamical regimes.
Since such critical transitions disrupt the systems' intended or desired functionality, it is crucial to understand the underlying mechanisms, to identify precursors of such transitions and to reliably detect them in time series of suitable system observables to enable forecasts.
This review critically assesses the various steps of investigation involved in time-series-analysis-based detection of critical transitions in real-world non-autonomous systems: from the data recording to evaluating the reliability of offline and online detections.
It will highlight pros and cons to stimulate further developments, which would be necessary to advance understanding and forecasting nonlinear behavior such as critical transitions in complex systems.
\end{abstract}

\maketitle

\begin{quotation}
Critical transitions are apparent sudden shifts in the states of complex dynamical systems and are often followed by extreme events that may entail loss of life and/or materials.
Reliable detection and forecasting of critical transitions would allow for developing adaptation and/or mitigation strategies, which is thus of utmost importance in many scientific fields.
Although we still lack a general understanding of the mechanisms underlying critical transitions, there is plenty of observational data which might qualify for a time-series-analysis-based detection of transitions in real-world non-autonomous systems.
Successful detection and forecasting of critical transitions requires not only knowledge of the various concepts and methodologies involved, but also an assessment of possible sources of error. 
This is the subject of the present review.
\end{quotation}

\section{Introduction}
%
Complex dynamical systems, ranging from the brain and the heart to ecosystems, financial markets and the earth and climate systems can show sudden transitions to very different and, at times, even disastrous dynamical regimes~\cite{Albeverio2006,Basher2006,Willoughby2007,lenton2011,ansmann2013,Meron2015,trefois2015,Intrieri2016,Bok2018,Sapsis2018,Hanifi2020,Morozov2020,Scheffer2020,Rundle2021,Bletery2023,Cerini2023,Domeisen2023,Lehnertz2023b,Flores2024,Hardebeck2024}.
Typical examples include epileptic seizures and cardiac arrest, population extinctions, market crashes, mass panics, wars,  earthquakes, tsunamis, extreme weather events, rogue waves, or large-scale blackouts in power supply networks. 
Observations and model simulations indicate that such critical transitions often occur when changing conditions pass a critical or tipping point~\cite{horsthemke1984,kuehn2011,ashwin2012,Feudel2023}.
Both reliable detection and forecasting of critical transitions are of utmost importance in many scientific fields. 
If critical catastrophic transitions occur in an unexpected way, they often do not allow for developing adaptation and/or mitigation strategies. 
It is thus crucial to understand the mechanisms underlying critical transitions, to identify early-warning indicators of such transitions, and to reliably detect them in time series of suitable system observables to enable forecasts.

The real-world systems referred to above can be considered as non-autonomous dynamical systems~\cite{Sell1967a,Kloeden2011}, for which the evolution of the state variable ${\bf x} \in \R^d$ is governed by
\begin{equation}
\frac{\dx {\bf x}(t)}{\dx t} = f(t,{\bf x}(t),\beta(t)), \qquad \frac{\partial f}{\partial t} \neq 0.
\label{eq:NADS}
\end{equation}
\revise{Such systems are usually subject to time-dependent forcing as well as to additive and/or multiplicative dynamical noise}, 
which --~together with the various possible transition mechanisms (bifurcation-, noise- , rate-, or shock-induced tipping~\cite{horsthemke1984,kuehn2011,ashwin2012,Feudel2023,Pavithran2023})~-- poses particular challenges for a time-series-analysis-based detection of critical transitions.

The last years have seen a rapid increase of data- and theory-driven approaches that aim at gaining deeper insights into non-autonomous dynamics~\cite{clemson2014}, into the mechanisms underlying critical transitions and at their detection and forecasting. 
Among others, methods based on \revise{nonlinear time-series analysis~\cite{kantz2003}, complex network theory~\cite{boccaletti2006,Zou2019}, or symbolic time-series representation~\cite{Amigo2010}} have been demonstrated to be useful for specific systems, however, the general applicability of these methods remains an issue to be investigated. 
Importantly, and with an eye on field applications, it is crucial to estimate the reliability and significance of indicators for critical transitions.

In this review, we summarize recent conceptual and methodological developments that aim at a time-series-analysis-based investigation of critical transitions in real-world non-autonomous systems.
We critically assess the various steps of investigation~--~from the experimental observation (Sect.~\ref{sect:experiment}) via means to cope with non-stationarity (Sect.~\ref{sect:nonstat}) and time-series analyses (Sect.~\ref{sect:tsa}) to the offline detection of transitions (Sect.~\ref{sect:detect}) and (statistical) means to evaluate their reliability (Sect.~\ref{sect:reliab_off})~--~and highlight areas that are under active investigation and that promise to provide new insights.
Eventually, we briefly discuss possibilities for time-series-analysis-based online detections of critical transitions that could contribute to further increase reliability of forecasts.

\section{Experimental observation of the dynamics of real-world non-autonomous systems}
\label{sect:experiment}
For many real-world non-autonomous dynamical systems the governing equations of motion are unknown, and quite often one lacks detailed knowledge about the mechanisms that may underlie transitions. 
In such cases, one typically resorts to the analysis of time series of suitable observables $\left\{s(t) = s({\bf x}(t)\right\}$ with the measurement function $s:\R^d \rightarrow \R$ and the discretized time $t=N\Delta t$ (with the number of data points $N$, and $\Delta t$ is the sampling interval).
This approach is, however, associated with some difficulties that one should be aware of in order to avoid misinterpretations.

Frequently, one has access to a few (or even to only one) dynamical variables, and one should consider in advance whether these are ``good'' observables~\cite{Kwasniok2004,Mignan2014,Mignan2021}. 

\begin{figure}[h]
\begin{centering}
 	\includegraphics[width=.75\columnwidth]{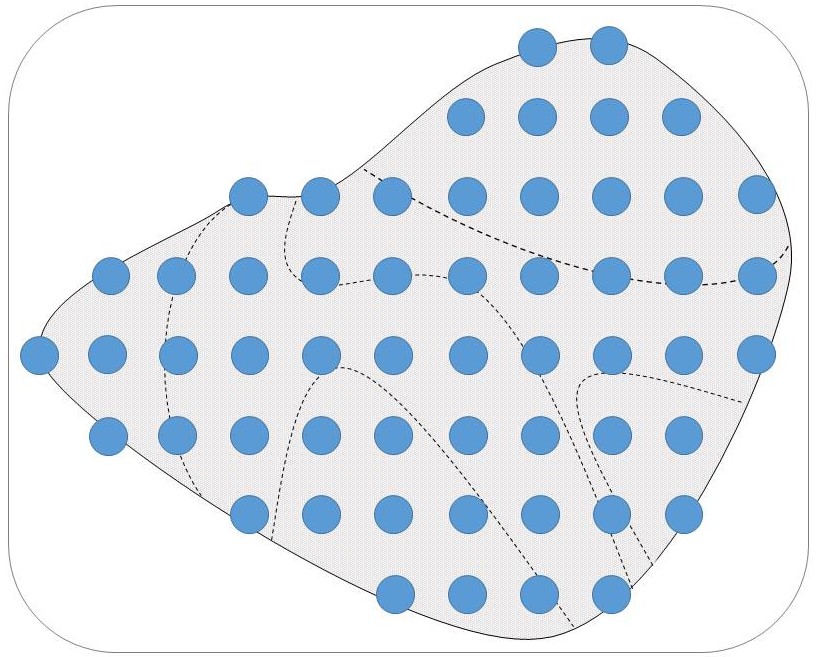}
	\end{centering}
	\caption{Sampling the dynamics of a spatially extended system --~for which the actual division into subsystems (black dotted lines) is not known a priori~-- is often performed with a grid-like arrangement of sensors (blue circles). 
	An associated problem is spatial oversampling whereby several sensors record the dynamics of a given subsystem. Such redundant and mostly noisy information might be hard to distinguish from synchronization or spreading phenomena. 
	Another problem is related to sensors that were happened to be placed such that they capture the dynamics of neighboring subsystems. 
	Disentangling superimposed dynamics is hard without having the necessary a priori information about the subsystems.}
	\label{fig:fig1}
\end{figure}

Lacking detailed knowledge about the system under investigation, one is faced with the problem of how to sufficiently sample the system dynamics in time and space.
This requires choosing a priori various recording parameters such as number and placement of sensors, the sampling interval $\Delta t$, filter and amplifier settings, and precision of analogue-to-digital conversion.
The well-known Nyquist–Shannon sampling theorem~\cite{Nyquist1928b,Shannon1949,Jerri1977} may have only limited significance, particularly if the full range of dynamical behaviors and their temporal and spatial scales are not known a priori. 
If the system contains nonlinearities, band-limiting possibly chaotic dynamics using inappropriate filters may lead to severe misinterpretations~\cite{kantz2003}.
Lacking detailed knowledge about the system's spatial scales and about how --~if at all~-- to decompose it into subsystems, one typically resorts to a grid-like arrangement of sensors (Fig.\ref{fig:fig1}).
In case of complex coupling topologies between subsystems, such a placement can lead to severe misinterpretations~\cite{bialonski2010,porz2014}.

A widely used approach to study a system's dynamics is probing, i.e., investigating the system's response (relaxation dynamics) to small perturbations. 
Such an approach, however, might not be feasible for any system and more importantly, it involves a high level of risk of inducing a transition to a new state with possibly undesirable properties --~you may not know in advance how big a small perturbation actually is. 
From the point-of-view of time-series analysis, relaxation dynamics are typically reflected in noisy, short-lasting transient signals that require dedicated denoising techniques~\cite{effern2000d} or multiple realizations~\cite{janosi1994,dhamala2001,andrzejak2006b,leski2008,komalapriya2008,wagner2010a,martini2011,Ma2014b,Wang2016} that may also be used to enhance time resolution by replacing the temporal average with an ensemble average. 
The latter, however, implicitly assumes stationarity of the system (at least on the considered timescale) and comparability of  relaxation dynamics and of transitions.

An alternative to probing is observing a system over timescales long enough to capture all relevant aspects of its dynamics (e.g. ``normal'' dynamics, critical and non-critical transitions).
For real-world non-autonomous dynamical systems, however, this approach requires dedicated methods to cope with the inherently nonstationary dynamics (see Sect.~\ref{sect:nonstat}), with possible very long transients~\cite{kaneko1990,tel2008,Morozov2020,MeyerOrtmanns2023}, or even with non-convergent (non-asymptotic) behaviors~\cite{ansmann2016,Nandan2023}.
It should be determined in advance whether continuous data collection is required or whether collecting snippets of data suffices for an adequate characterization of the system's dynamics.
Likewise, it is advisable to keep the aforementioned recording parameters constant throughout the observation time as these could easily represent time-dependent confounders.

\section{Coping with non-stationarity}
\label{sect:nonstat}
Non-autonomous dynamical systems are non-stationary, non-ergodic systems that do not obey \revise{Birkhoff’s 
ergodic theorem equaling time and ensemble averages}~\cite{Boltzmann1877,Birkhoff1931}.
This property is usually regarded as an obstacle as it poses special challenges for time-series-analysis methods.

A (stochastic) process is called strictly (or strongly) stationary if the distribution of its states over an ensemble of realizations of that process does not depend on time~\cite{priestley1988}.
This implies constancy of all statistical moments and all joint statistical moments.
Within the scope of field applications, stationarity of a system can only be evaluated with respect to the observation time.
The vast majority of time-series-analysis methods (see Sect.~\ref{sect:tsa}) require the system under investigation to be at least approximately stationary (constancy of first- and second-order statistical moments) in order to allow for a robust and reliable characterization.

\begin{figure*}[t]
\begin{centering}
 	\includegraphics[width=0.9\textwidth]{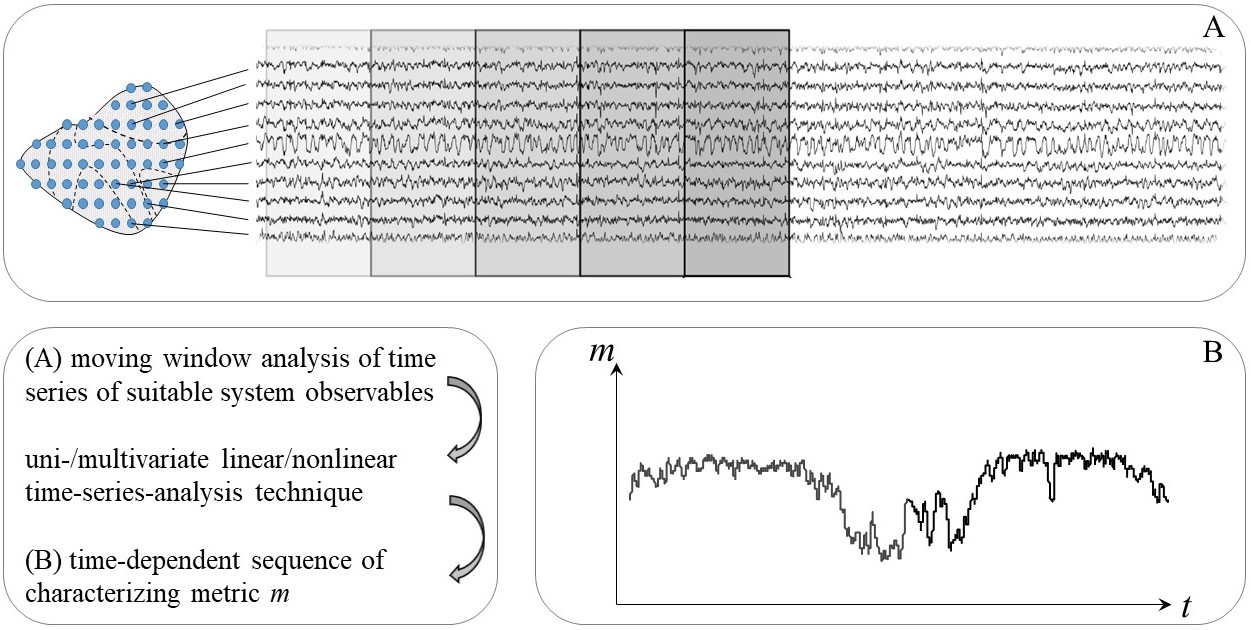}
	\end{centering}
	\caption{A frequently used approach to investigate the dynamics of a non-stationary system is to estimate a characterizing metric $m$ (cf. Sect.~\ref{sect:tsa}) using the data of successive segments (or windows; marked in gray) of time series of system observables. 
	During each segment, the system is assumed to be (approximately) stationary (moving-window analysis).
	The resulting time-dependent sequence $m(t_{\rm w})$, where $t_{\rm w}$ denotes the sequence's temporal resolution,  is then used to detect transitions as well as to identify indicators for the latter.}
	\label{fig:fig2}
\end{figure*}

The most common way of dealing with a non-stationary system is to cut the time series into successive (non\nobreakdash-)overlapping segments (or windows~\cite{Prabhu2014}) during which the system's dynamics can be regarded as approximately stationary (sliding or rolling or moving-window analysis)~\cite{Kitagawa1987,Dahlhaus1997,huang1998,Verdes2001,Kantelhardt2002,Fukuda2004,Wu2007b,clemson2014,lehnertz2017,Podobnik2008,Rhif2019}.
A characterizing metric is then calculated for each data window by employing some time-series-analysis method (see Fig.~\ref{fig:fig2} and Sect.~\ref{sect:tsa}).

Instead of identifying (approximately) stationary data segments, one might also consider non-stationarity as an interesting aspect of the dynamics.
Methods that allow tracing non-stationarity are based e.g. on the concept of overembedding~\cite{hegger2000,Verdes2006,DeDomenico2010} that treats time-varying control parameters as independent dynamical variables.
Other methods make use of recurrence quantification analysis~\cite{Manuca1996,kennel1997,Gao2001,rieke2002,rieke2004,Facchini2005,Chen2012b}, of state-space-based cross-predictions~\cite{schreiber1997}, or of symbolic data compression~\cite{Kennel2000}, to name just a few.

Regardless of the perspective taken, all approaches require some statistical test~\cite{priestley1988,Robinson1994,witt1998} to assess the quality of differentiation/characterization.

\section{Time-series-analysis}
\label{sect:tsa}
Time-series analysis comprises methods for analyzing a single or multiple sequences of observations collected over a certain period of time in order to extract meaningful statistics and other characteristics of the data.
A plethora of linear and nonlinear time-series-analysis methods is available that can be used to detect and characterize --~in a data-driven way~-- different aspects and properties of a system's dynamics~\cite{champeney1973,brody1981,haykin1983,grassberger1991a,abarbanel1993,honerkamp1993,kaplan-glass1995,Percival2000,pikovsky2001,daw2003,kantz2003,reinsel2003,Keller2005,lutkepohl2005,hlavackova2007,marwan2007,lacasa2008,Anderson2011,friedrich2011,VonToussaint2011,Bradley2015,webber2015,stankovski2017,tabar2019book,Zou2019,Hamilton2020,Edinburgh2021,Datseris2022,Fokianos2022,Nikakhtar2023,Tabar2024}.
Some methods may also be used for forecasting (or anticipation) purposes.

Analysis methods can be classified into two main groups. 

With univariate methods, the above-mentioned applications are limited to a single time series.
Exemplary applications include estimating statistical moments (e.g., mean, standard deviation, skewness, kurtosis), characteristics of the auto-correlation function (e.g., decay time) or the power spectrum (e.g., spectral power in some frequency band), Kramers-Moyal coefficients (e.g., drift and diffusion), \revise{or state-space-based quantities such as dimensions, entropies, stability indicators like Lyapunov exponents, or characteristics derived from recurrence quantification analysis.
The latter quantities require an appropriate reconstruction of the dynamics in state-space from a single (or only a few) time series~\cite{packard1980,takens1981,sauer1991,Casdagli1991,kennel1992,Kugiumtzis1996,cao97,Cellucci2003,vlachos2010,Kraemer2021}.}

With bivariate (or, in general, multivariate) methods, investigations mostly concentrate on characterizing interdependences between pairs (or, in general, n-tuples) of time series. 
Exemplary applications include estimating characteristics of the cross-correlation function (e.g., interaction delay) or the cross-spectrum (e.g., coherence function), properties of interactions (strength, direction, functional form) and of synchronization phenomena, or characteristics derived from network theory.

Since there is no one-fits-all time-series-analysis method for all types of data, the choice of a specific method depends on the system under investigation and on the research question being asked.
In addition, since one is interested in the physical mechanism(s) underlying the observed dynamics (or properties thereof), the choice of the method needs to be justified by a hypothesis about the appropriate data model.
Given that almost all time-series analysis methods reduce the information content of some time series to some characteristic number or metric, interpretability of the latter is only possible if it has a specific meaning within the model framework.
Only in that case does the reduced information increase our knowledge about the system under investigation.
If the data to be analyzed does not pertain to the appropriate model class, the chosen metric may not make much sense, even if its numerical value can be calculated for the given time series using a numerical algorithm.

For the majority of time-series-analysis methods, a reduction of the information content of some time series involves time-, space-, or ensemble-averages or combinations thereof in order to achieve a sufficient accuracy and robustness against noise contaminations. 
\revise{It is often assumed} that the more data enters some algorithm the higher is the method's accuracy and thus the reliability of the chosen metric.
However, many time-series-analysis methods, or rather the concepts/theories from which they originate, (implicitly) assume the system under investigation to be stationary.
As already mentioned above, investigations of non-autonomous dynamical systems often rely on a moving-window analysis to account for non-stationarity. 
A shorter analysis window is usually associated with approximate stationarity, but one needs to take into account that the decreased amount of data entering some analysis method may decrease the reliability of a characterizing metric. 
Thus, the choice of a window length $N_{\rm{w}} \ll N$ is often a compromise between the required statistical accuracy for the estimation of a metric and approximate stationarity within a window length.

Calculating a metric using a time-series-analysis method is usually accompanied by the choice of algorithmic parameters. 
Depending on the research question, algorithmic parameters are often tuned to achieve an optimum characterization of the system's dynamics. 
For an investigation of long-term observations and using a moving-window analysis, however, tuning parameters for each window might not be advisable.
The resulting (time-dependent) sequence of a metric's numerical values could then simply reflect this tuning-per-window which can lead to severe misinterpretations.
It might thus be advisable to fix all necessary pre-processing steps and keep a priori best-chosen algorithmic parameters constant. 

\section{Detecting transitions}
\label{sect:detect}
A number of metrics have been proposed in recent years to detect critical transitions based on bifurcations and the associated phenomenon of critical slowing down, i.e., an increasingly slow recovery from small perturbations~\cite{dakos2008,scheffer2009,scheffer2012,dai2012,lenton2012}.
With approaching the bifurcation (tipping point), the time needed to recover from perturbations becomes longer and hence the system dynamics becomes more correlated with its past, leading to an increase in the lag-1 autocorrelation and to spectral reddening~\cite{Bury2020}.
Since perturbations accumulate, one observes an increase in the size of the fluctuations~\cite{kubo1966} as assessed with variance or other higher-order statistical moments.
These classic statistical indicators have been used to investigate critical slowing in various contexts~\cite{CotillaSanchez2012,dakos2012,leemput2014,meisel2015b,Perry2015b,Nazarimehr2020,Brookes2021,George2021,Southall2021,Dablander2022,Deb2022,Ditlevsen2023,Mathevet2024} and were claimed to be generic.
They continue, however, to be critically discussed from various perspectives~\cite{ditlevsen2010,boettiger2012,boettiger2013c,boettiger2013b,Kefi2013,guttal2013,dakos2015,dai2015a,diks2015,wagner2015,zhang2015a,Guttal2016,Gsell2016,milanowski2016,Dutta2018,wen2018,romano2018,arumugam2019,clements2019,gatfaoui2019,jager2019,wilkat2019,Marconi2020,vanderBolt2021,Lapeyrolerie2023,OBrien2023,Proverbio2023}.

The quest for more robust and possibly more widely applicable indicators that may be sensitive to critical transitions other than those related to bifurcations 
\revise{--~ such as boundary crisis~\cite{grebogi1983,Osinga2000}, blowout bifurcations~\cite{ott1994a,Zhang2020}, saddle escape~\cite{kuehn2015b}, oscillation death~\cite{koseska2013,Zou2021}, or explosive synchronization~\cite{boccaletti2016b,Kuehn2021}~--}
has led to the development of metrics that are based on other time-series-analysis concepts. 
Among others,  these include synchronization~\cite{mormann2003a,mormann2005,winterhalder2006c,kuhlmann2010}, symbolic dynamics~\cite{Ray2004,Chin2005,lehnertz-dickten2015}, complex networks~\cite{liu2015,Peng2019,rings2019trace,rings2019precursors,Ludescher2021,Lehnertz2023b}, algebraic topology~\cite{Mittal2017,SyedMusa2021,Ghil2023}, autoregressive modeling~\cite{Giacomini2009,Ghalati2019}, entropy concepts~\cite{Liu2012,Meng2020,Pavithran2021,Tirabassi2023,Deb2024},  reconstruction of Jacobian matrices~\cite{Barter2021}, (nonstationary) dynamical modeling~\cite{Franzke2013b,Kwasniok2013,Kwasniok2015,Kwasniok2018,Din2018}, or the Kramers-Moyal expansion~\cite{lehnertz2018,tabar2019book,Arani2021,Hessler2022,Hessler2023}.

All these time-series-analysis techniques have in common that, when investigating real-world non-autonomous dynamical systems, moving-window analyses of some system observable(s) yield time-dependent sequence(s) of a characterizing metric $\left\{m(t_{\rm{w}}), m(2t_{\rm{w}}), m(3t_{\rm{w}}), \dots\right\}$.
These sequences have a temporal resolution $t_{\rm{w}}$ that is given by the window length $N_{\rm{w}}$ and can subsequently be used to detect transitions as well as to identify indicators for the latter.

\begin{figure}[h]
\begin{centering}
 	\includegraphics[width=.9\columnwidth]{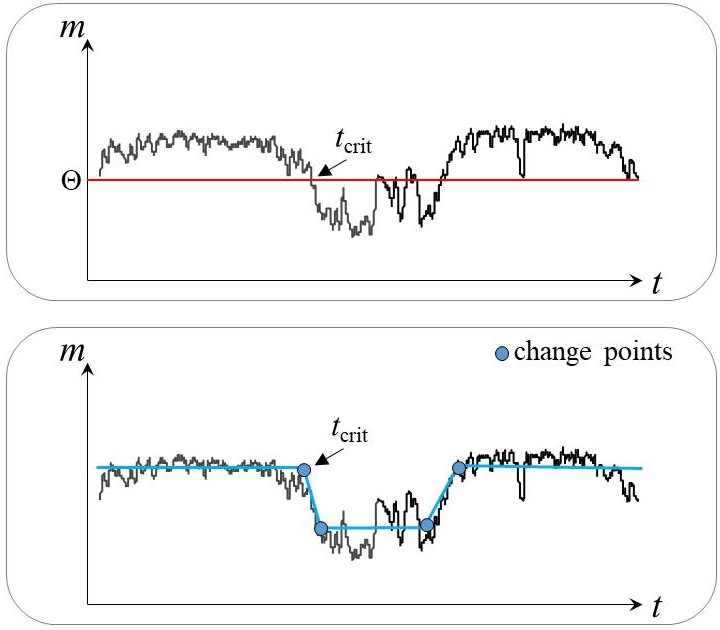}
	\end{centering}
	\caption{Illustration of techniques to detect transitions. 
	With threshold crossing (top), the onset of a transition is related to the first time point $t_{\rm{crit}}$ when a characterizing metric $m$ crosses a pre-defined threshold $\Theta$ (either from above or from below). 
	With change point detection (bottom), specific points in time (blue dots) are identified where the sequence of metric $m$ exhibits a significant change either in its deterministic content or underlying stochastic distribution. The onset of a transition can be related, for example, to the first identified change point.}
	\label{fig:fig3}
\end{figure}

There are two major techniques in use to detect transitions, namely threshold crossing and change point detection (see Fig.\ref{fig:fig3}).
The latter is sometime also referred to as anomaly detection. 
Both techniques are often employed as offline methods that retrospectively detect changes when the data has already been gathered and using information on transitions that have taken place in the past.
When employed as online methods, they aim to detect changes as soon as they occur in a real-time setting.
This may be achieved with machine learning techniques~\cite{Lim2020,Kong2021,Ray2021,Patel2021,Patel2023,Koeglmayr2024}
or Bayesian inference~\cite{Shcherbakov2019,Hessler2022}.

\subsection*{Threshold crossing}
The onset of a transition is given by the first timepoint $t_{\rm{crit}}$ when the metric $m$ exceeds or falls below some pre-defined threshold $\Theta$~\cite{lehnertz1998,Streeter2013,bialonski2015,zhang2015a,Grziwotz2023}. 
The latter may be defined as the mean plus or minus two standard deviations estimated from preceding data (two-sigma rule~\cite{Benjamin2018}).
Note, however, that in driven non-stationary systems statistical moments of a sequence of some metric may undergo fluctuations on timescales much longer than the one for which the system's dynamics has been regarded as approximately stationary~\cite{Lehnertz2021a} (see Sect.~\ref{sect:nonstat}).
In such a case it might be advisable to replace the fixed threshold by an adaptive one~\cite{Dahlhaus2001}.
Since the rather simple threshold-crossing-based criterion might be prone to isolated outliers, one can also consider the onset of a transition as the first timepoint for which multiple consecutive data points $\left\{ m(t_{\rm{crit}}), m(t_{\rm{crit}}+t_{\rm{w}}), m(t_{\rm{crit}}+2t_{\rm{w}}), \dots \right\}$ fulfill the threshold criterion.
Indeed, one can expect the number of consecutive data points fulfilling the threshold criterion to increase as the system approaches a critical transition.
If assumptions can be made about how a metric changes during a critical transition (e.g. linear increase or decrease), one may eventually establish the presence of trends (e.g., using linear regression) in order to further substantiate the evidence of a critical transition. 

\subsection*{Change point detection}
This technique aims to identify specific points in time where the sequence of metric $m$ exhibits a significant change either in its deterministic content or underlying stochastic distribution~\cite{Carlstein1988,Basseville1993,Aminikhanghahi2017,Cabrieto2017,Cabrieto2018,Truong2020,Bagniewski2021,Bury2021,Gilarranz2022}. 
In general, change point detection concerns both detecting whether or not a change has occurred, or whether several changes might have occurred, and identifying the times of any such changes.
Depending on the application, this may involve measuring the statistical (dis\nobreakdash-)similarity between two windows of data to either side of a purported change point or \revise{identifying changes in the mean, variance, autocorrelation, or spectral density of the sequence~\cite{Truong2020,DeRyck2021}.} 
Note that the latter also requires the choice of a suitable and a priori defined window.
In addition, the same restrictions apply in the case of driven non-stationary systems as with threshold crossing.

\section{Testing the reliability of offline detections}
\label{sect:reliab_off}
Consider an offline time-series-analysis-based detection of critical transitions in some real-world non-autonomous dynamical system for which we know the exact onset times of critical transitions. 
The applied analyses techniques are assumed to be sensitive enough to detect and characterize critical transitions, based on prior investigations (e.g., with toy models).
We observe additional changes in the temporal sequence of the employed metric that would qualify as indicative of critical transitions, but we know a posteriori that the system has solely undergone transitions to some non-critical states --~we refer to these as non-critical transitions. 
\revise{Such a situation can occur, for example, when analyzing brain dynamics and employing time-series-analysis techniques that can not (or only insufficiently) distinguish between a transition to a pre-seizure state (critical transition) and transitions to behavioral states like sleep or wakefulness.}
We consider all other changes as false detections.
\begin{figure}[h]
\begin{centering}
 	\includegraphics[width=.9\columnwidth]{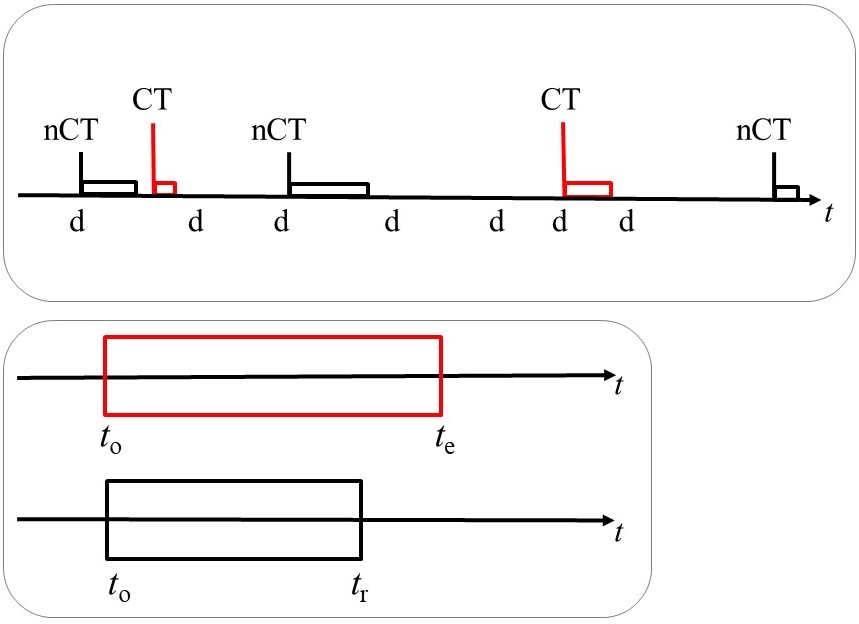}
	\end{centering}
	\caption{Top: exemplary outcome of an offline detection of the onsets of critical transitions (CT, red). Onsets of non-critical transitions (nCT) are marked black. Detections are labeled with the letter ``d''.
	 Bottom: the duration of a critical ($\Delta_{\rm CT}$, red) and a non-critical ($\Delta_{\rm nCT}$, black) transition can be defined as differences between the respective ending and onset times.
	Onset times are denoted by $t_{\rm o}$; for the critical transition, the ending of the transitory phase coincides with the occurrence of the extreme event at $t_{\rm e}$ (note that for $\Delta_{\rm CT} \to 0$, the transition would be classified as abrupt and there are no early warning signs); 
	for the non-critical transition, the ending of the transitory phase coincides with the return of the system to its \textit{normal} dynamics at $t_{\rm r}$.}
	\label{fig:fig4}
\end{figure}

For the (quite typical) detection example shown in the upper part of Figure~\ref{fig:fig4}, the following interpretations can be made: 
\begin{itemize}
\item since the method detected only half of the critical transitions, it appears to be not very sensitive; 
\item given that the method detected two out of three non-critical transitions, it might be more sensitive to the latter type of transitions;
\item there are more false than true detections, so the method appears to be not very specific;
\item there are two detections that precede the onsets of the critical transitions, however, since one detection coincides with the onset of a non-critical transition the method appears to have only a limited predictive significance;
\item since there are three detections that follow the offsets of critical and non-critical transition, one might speculate about the method's capability to detect the offset of such transitions.
\end{itemize}

This example clearly points to the need to check reliability and quality of offline time-series-analysis-based methods for the detection of critical transitions\cite{Boettiger2012b}.
For this purpose, various statistical tests can (and should) be applied in order to avoid drawing false conclusions and to avoid wishful thinking.

Tests range from simple to advanced and require knowledge about the occurrence times of (non\nobreakdash-)critical transitions (or events) as well as about characteristics of the respective transitory phases that may differ from event to event (in order to simplify things, one could treat false detections as non-critical transitions). 
Depending on the application, these characteristics include e.g. the durations of transitory phases ($\Delta_{\rm CT}$ and $\Delta_{\rm nCT}$; see lower part of Figure~\ref{fig:fig4}), the maximum deflection from the threshold, or characteristics of a trend function~\cite{Bunde2022}.  

\subsection*{Distinguishability of non-critical and critical transitory phases}
A simple though not very reliable way to assess detection performance of offline methods is to use statistical approaches~\cite{Daniel2000} to verify distinguishability of non-critical and critical transitory phases by comparing the distributions --~or moments thereof~-- of their characteristics (durations, maximum deflections, etc.).
The equality of distributions can be tested nonparametrically~\cite{Kolmogorov1933} or rank-based~\cite{Mann1947}.
For Gaussian-distributed data, equality of means and standard deviations can be evaluated with the Student's t-test and the F-test, respectively.
For non-Gaussian-distributed data, these are the median test or the Mann-Whitney U test and the Conover squared ranks test. 
It should be noted, however, that the tests assume that the data were taken independently (from some distribution).
This implies no memory, no dynamics, time is not important, and stationarity of the investigated system.

\subsection*{Sensitivity analysis}
Another simple way to assess detection performance of offline methods is to count the number of correct detections of critical transitions.
This approach allows drawing some conclusions about a method's sensitivity but it does say anything about a method's specificity.
Sensitivity refers to the probability of a positive test result (``is critical''), conditioned on each detection truly being positive.
Specificity refers to the probability of a negative test result (``is non-critical''), conditioned on each detection truly being negative.
Also, this approach does not take into account false detections (see below), and one quite often finds studies that report on sensitivity of a detection method only.
Note that a detection method can easily be tuned through in-sample optimization to achieve 100\,\% sensitivity when ignoring its specificity. 
Lacking relevant information and, more importantly, considering the fact that critical transitions are typically rare events, this simple way is not very robust and largely unsuited for analyses of empirical data.

\subsection*{ROC analysis}
\begin{figure}[h]
\begin{centering}
 	\includegraphics[width=.65\columnwidth]{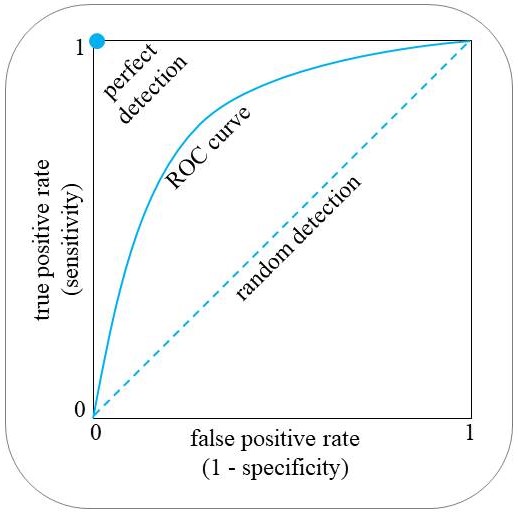}
	\end{centering}
	\caption{Performance of a detection method can be evaluated on the basis of a receiver operating characteristic (ROC) curve that relates true positive rates to false positive rates. 
	The area under the curve (AUC) can be used to quantify the detection method's performance (an ideal detection method has an AUC of 1).}
	\label{fig:fig5}
\end{figure}

A more robust way to assess detection performance is the Receiver Operating Characteristics (ROC) analysis~\cite{Yerushalmy1947,Fawcett2006}.
In addition to counting the number of correct (true positive) and missing (true negative) detections of critical transitions as estimates for sensitivity and specificity, this approach also takes into account false detections.  
A false positive detection refers to a non-critical transition that was erroneously detected as critical transition (type I error).
A false negative detection refers to a critical transition that was erroneously detected as non-critical transition (type II error).
One then estimates the method's sensitivity (or the true positive rate; TPR) as well as $1-$specificity (or the false positive rate; FPR) at each setting of the threshold value $\Theta$ to obtain the so called ROC curve~\cite{egan1975} (see Fig.~\ref{fig:fig5}). 
In this ROC space, the best possible detection method would yield a point in the upper left corner (or lower right depending on interpretation and change of metric $m$), representing 100\,\% sensitivity (no false negatives) and 100\,\% specificity (no false positives). 
In contrast, a detection method working at random would give a point along the diagonal from the bottom left to the top right corners (regardless of the positive and negative base rates; cf. random guessing).

An often used reduction of the ROC Curve to a single number is calculating the so called area under the ROC curve (AUC).
Depending on application, AUC can be defined as the value of the area under the whole curve which yields 1.0 for the best possible detection method and 0.5 for a random detection~\cite{mormann2005}. 
Alternatively, AUC can be defined as the (absolute) value of the area between the ROC curve and the diagonal which yields values between 0.0 (random detection) and 0.5 (best detection).

High AUC values for various detection methods continue to be reported from different disciplines~\cite{Drake2013,zhang2015a,brinkmann2016,Cavaliere2016,kuhlmann2018b,Yang2018b,gatfaoui2019,Brett2020,Bury2021,Southall2021,Tredennick2022,Dylewsky2023,GomezNava2023}.
Such values, however, can give a false impression about a method's genuine detection reliability, particularly for imbalanced probability distributions for true and false positive detections, which is the case for critical transitions that are typically rare events.
This imbalance can lead to a bias towards a predominant number of detections of non-critical transitions and can render an interpretation of AUC values oversimplistic and even misleading~\cite{Weiss2004,mormann2005,BenBouallegue2022,West2023}.
Such misinterpretations can be avoided with the help of surrogate-assisted ROC analyses that were introduced in the field of prediction of epileptic seizures~\cite{mormann2007,kuhlmann2018} in the context of detection of critical transitions (here: detection of a pre-seizure state).

\subsection*{Surrogate-guided ROC analysis}
A simple and straightforward ansatz consists of comparing a method's detection performance against the one of a random detector~\cite{winterhalder2003,schelter2006,feldwisch2011b,Mader2014}.
To do so, artificial detections are generated at random times (e.g., Poisson-distributed) but using the same rate as that of the method under investigation. 
This easy and computationally fast approach might require adjustments in case of (near-)periodic occurrences of transitions and it might not be generally applicable.
A more advanced though more time-intensive ansatz makes use of null-hypothesis-based resampling or bootstrapping methods~\cite{efron1982,efron1998}, also known as surrogate techniques in time series~\cite{schreiber2000a,lucio2012,lancaster2018} and network analysis~\cite{ansmann2011,ansmann2012,bialonski2012,laut2016,wiedermann2016,stahn2017,Chorozoglou2019b}.
One such technique~\cite{andrzejak2003} consists of a random shuffling of the time intervals between successive (true) detections of critical transitions to generate artificial detection timepoints.
A related technique~\cite{kreuz2004} leaves the (true) detection timepoints unchanged and instead performs a constrained randomization of the temporal sequence of a characterizing metric using e.g. simulated annealing~\cite{Bertsimas1993}.
Both these techniques require formulating an appropriate null-hypothesis as well as a robust (non-parametric) test statistic to assess the statistical significance of the results obtained.
The number of independent surrogates $N_{\rm surr}$ determines the nominal size of the test statistic (i.e., the probability $\alpha$ of making a type I error; $1/N_{\rm surr} \leq \alpha$).
The performance (as assessed with AUC) of the detection method under investigation is then compared with the performances of a surrogate ensemble. 
If the method's performance exceeds to performances of the surrogates, then the method can be considered --~with greater confidence~-- as performing better than chance.
Reporting the ``number of sigmas'' can be misleading since it is accompanied by strict assumptions about the underlying distributions of performance values. 
Surrogate-based approaches can account for a possibly non-random occurrence of critical transitions which might be encountered with driven non-stationary systems. 

\subsection*{Time series surrogates}
Null-hypothesis-based surrogate techniques are also being used to generate ensembles of time series that share relevant statistical properties with the original time series of some suitable observable except for properties that are assumed to be indicative of a critical transition. 
As an example, we mention trends of indicators for critical slowing down such as variance and lag-1 autocorrelation. 
Techniques to generate such surrogate time series include random shuffling, autoregressive model-based methods, or iterative amplitude-adjusted Fourier transform~\cite{schreiber2000a} and are often used to evaluate the significance of the aforementioned indicators~\cite{dakos2012,kramer2012b,George2021,Boettner2022,Chen2022b,Pal2022,Bochow2023}.
Techniques require formulating an appropriate null-hypothesis as well as a robust (non-parametric) test statistic to assess the statistical significance of the results obtained.
The number of independent surrogate time series determines the nominal size of the test statistic.\\

Before closing this section, it is important to recall the following caveats when interpreting an offline detection method's reliability evaluated with the aforementioned tests\cite{Shmueli2010}:

\begin{itemize}
\item statistical tests can neither prove the correctness of an offline detection method's reliability nor can they prove the existence of a critical transition;
\item a rejection of a null hypothesis only provides a necessary but not a sufficient condition for a method to indicate reliable offline detection of a critical transition;
\item an acceptance of a null hypothesis does not indicate its correctness;
\item whenever a null hypothesis is rejected, it is always very important to keep in mind that the complementary hypothesis is typically very comprehensive and might include many different reasons that are possibly responsible for this rejection;
\item consider including other (statistical) properties of your data and detection method into the null hypothesis.
\end{itemize}

\section{Prospects for time-series-analysis-based online detections of critical transitions}
\label{sect:online}
So far, only a few studies reported on above-chance-level online detections of critical transitions in real-world non-autonomous systems from empirical data~\cite{cook2013,Ludescher2023}.
Findings suggest that it is possible to forecast the occurrence of a \revise{recurrent} extreme event in some cases, however, predicting exact timing and magnitude of an event is much more difficult. 
Forecasts should thus be probabilistic~\cite{Gneiting2014} instead of aiming at fully deterministic predictions.

When judging the reliability of probabilistic forecasts (or forecast skill) of rare \revise{and recurrent} extreme events, the statistical tests discussed in Sect.~\ref{sect:reliab_off} can be oversimplistic or even misleading~\cite{Tashman2000,BenBouallegue2022}.
Proper scoring rules and strictly proper scores~\cite{Gneiting2007,Gneiting2011,lerch2017,Gneiting2023} and uncertainty quantification~\cite{Berger2019b} appear to be more suitable, however, their use for time-series-analysis-based online detections of critical transitions requires further conceptual and methodological developments.

\section{Conclusions}
The time-series-analysis-based detection of critical transitions in real-world non-autonomous systems is a young area of research
that carries the potential to advance understanding and reliable forecasting of such nonlinear behavior.
In this paper, I provided an overview of and critically assessed the various involved steps of investigation: from the data recording via coping with non-stationarity when analyzing time series of system observables to evaluating the reliability of offline and online detections.
I highlighted pros and cons to stimulate further developments that require a close interlinking of theory and experiment.

Time-series analysis methods in use appear to be suitable to allow detection and sufficient characterization of bifurcation-induced critical transitions in low-dimensional systems.
Nevertheless, their suitability for these types of transition in high-dimensional systems as well as for other transition mechanisms remains to be shown.
Is is conceivable that the latter require novel developments also in terms of robustness, appropriateness for non-stationary systems, and --~ideally~-- general interpretability.
Likewise, the same applies to methods to test the reliability of detections.

Extreme events occur in a variety of natural, technical, and societal systems and their frequency appears to be increasing.
Their catastrophic consequences and their low-probability, high-impact nature not only calls for improving our understanding of generating mechanisms but also to provide reliable early warnings to enable the development of adaptation and/or mitigation strategies.

\begin{acknowledgments}
The author acknowledges fruitful discussions with Manuel Adams, Timo Br\"ohl, Ulrike Feudel, and Thorsten Rings.
\end{acknowledgments}


\section*{AUTHOR DECLARATIONS}

\subsection*{Conflict of Interest}
The author has no conflicts to disclose.

\subsection*{Data availability}
Data sharing is not applicable to this article as no new data were created or analyzed in this study.



%

\end{document}